\documentclass[aps,prl,reprint,twocolumn,superscriptaddress,longbibliography,nofootinbib,floatfix,showpacs]{revtex4-1}
\usepackage{epsfig}
\usepackage{amsmath,amssymb,amsfonts}
\usepackage{hyperref}
\usepackage{mathrsfs}
\usepackage{bbm}
\usepackage{slashed}
\usepackage{graphicx}
\usepackage{verbatim}

\usepackage[usenames]{color}

\newcommand{\mcE}{\mathcal{E}}
\newcommand{\Eqref}[1]{Eq.~\eqref{#1}}
\newcommand{\be}{\begin{equation}}
\newcommand{\ee}{\end{equation}}
\newcommand{\bw}{\begin{widetext}}
\newcommand{\ew}{\end{widetext}}
\newcommand{\bi}{\begin{itemize}}
\newcommand{\ei}{\end{itemize}}
\newcommand{\bea}{\begin{eqnarray}}
\newcommand{\eea}{\end{eqnarray}}
\newcommand{\ud}{\mathrm{d}}
\newcommand{\critexp}{\beta}

\usepackage[T1]{fontenc} \usepackage[latin1]{inputenc}

\begin{document}

\title{Critical Schwinger pair production}

\author{Holger Gies}
\email[]{holger.gies@uni-jena.de}
\affiliation{Theoretisch-Physikalisches Institut, Abbe Center of Photonics, Friedrich-Schiller-Universit\"at Jena, 
Max-Wien-Platz 1, D-07743 Jena, Germany}   
\affiliation{Helmholtz-Institut Jena, Fr\"obelstieg 3, D-07743 Jena, Germany}
\author{Greger Torgrimsson}
\email[]{greger.torgrimsson@chalmers.se}
\affiliation{Department of Applied Physics, Chalmers University of Technology, SE-41296 Gothenburg, Sweden}

\begin{abstract}
We investigate Schwinger pair production in spatially inhomogeneous
electric backgrounds. A critical point for the onset of pair
production can be approached by fields that marginally provide sufficient electrostatic energy for an
off-shell long-range electron-positron fluctuation to become a real
pair. Close to this critical point, we observe features of
universality which are analogous to continuous phase transitions in
critical phenomena with the pair-production rate serving as an order
parameter: electric backgrounds can be subdivided into universality
classes and the onset of pair production exhibits characteristic
scaling laws. An appropriate design of the electric background field
can interpolate between power-law scaling, essential BKT-type scaling
and a power-law scaling with log corrections. The corresponding
critical exponents only depend on the large-scale features of the
electric background, whereas the microscopic details of the background
play the role of irrelevant perturbations not affecting criticality.
\end{abstract}

\maketitle

\section{Introduction}

Universality is an overarching concept in physics, signifying the
independence of general gross properties of a physical system of the
details of its microscopic realizations. Most prominently, critical
phenomena near a continuous phase transition reveal a remarkably high
degree of universality, such that different systems consisting
microscopically of rather different building blocks exhibit
quantitatively identical long-range behavior near the phase transition
\cite{Kadanoff:1971pc}. The quantification of universality by means of
fixed points is one of the great successes of the renormalization
group that provides a map from the microscopic details to the
effective long-range properties \cite{Wegner:1972ih}.

As a consequence, critical systems can be associated with universality
classes which are characterized by only a few properties such as the
symmetries of the order parameter, the dimensionality, and the number
and type of long-range degrees of freedom. It is therefore not
surprising that universality and a notion of criticality can also be
found beyond the realm of statistical physics. For instance, the onset
of black-hole formation shows a surprising insensitivity to the
initial data. Generically, the black hole mass as a function of a
single control parameter parametrizing the initial data scales
according to a power-law with the universal Choptuik exponent
\cite{Choptuik:1992jv,Gundlach:2007gc}. Whereas universality in
statistical physics is typically associated with the presence of
fluctuations on all scales, the example of gravitational collapse is
observed in a purely classical deterministic setting.

In the present work, we identify for the first time aspects of
universality in the phenomenon of Schwinger pair production in quantum
electrodynamics (QED). This sets a dual example as the
phenomenon of pair production in strong external fields can be derived
from the Dirac equation which -- despite its quantum mechanical
interpretation -- can be viewed as a classical deterministic field
equation. In fact, the first observation of this phenomenon relied on
this formulation \cite{Sauter:1931zz}, and so do many more modern
approaches at least indirectly
\cite{Brezin:1970xf,Gavrilov:1996pz,Kim:2000un,BialynickiBirula:1991tx,Ruf:2009zz,Hebenstreit:2013qxa}. On 
the other hand Schwinger pair production is also encoded in the photon
correlation functions derived from the full functional integral of
QED, as seen from the derivations of Euler, Heisenberg
\cite{Heisenberg:1935qt}, and Schwinger \cite{Schwinger:1951nm}. Again
many variants of this fluctuation-based descriptions exist
\cite{Dittrich:1985yb,Dunne:2009gi,Gies:2005bz}. The fact that both
descriptions are equivalent is a manifestation of the optical theorem
(for a recent discussion in the context of pair production, see
\cite{Dinu:2013gaa,Meuren:2014uia}).

In this work, we use the worldline formalism with background fields
\cite{Schmidt:1993rk}, as this method makes universality in the language of fluctuations of
electrons in spacetime most transparent.  We use the imaginary part of
the QED effective action $\Gamma$ as the order parameter for the onset
of criticality. It is related to the probability of vacuum decay via
$P=1-\exp(-2\,\text{Im }\Gamma[E])$ in the presence of an electric
field $E$; to lowest order, it is also related to the pair production
rate \cite{Nikishov:1970br,Cohen:2008wz}. The seminal
Schwinger formula
\be\label{Schwinger}
\text{Im }\Gamma=V_4\frac{(eE)^2}{8\pi^3}\sum\limits_{n=1}^\infty\frac{1}{n^2}\exp\left(-n\frac{\pi m^2}{eE}\right) 
\ee
exhibits no signature of criticality, as it assumes the presence of an
electric field being constant all throughout space and time. By
contrast, a critical point can arise for spatially inhomogeneous
fields, as can be read off from Nikishov's exact solution for the
electric field with the localized Sauter-profile $E(x)=\mcE\text{sech}^2
kx$ of inverse width $k$ \cite{Nikishov:1970br}. The order parameter for pair
production $\text{Im }\Gamma$ for such a spatial profile drops to zero at
\begin{equation}
e\!\int\! dx E(x) = \frac{2e\mcE}{k} \stackrel{!}{=}2 m\quad  \Rightarrow\,\, {\gamma_{\text{cr}}}:=\frac{k m}{e\mcE}=1. 
\label{eq:work}
\end{equation}
This equation has a simple meaning: the work done by the electric
field on a particle of charge $e$ propagating along the whole real
axis has to be equal to the rest mass of the particle--anti-particle
pair to be created. Translated into the language of fluctuations, a
virtual pair created at some spacetime point in such a field with
adiabaticity (Keldish) parameter $\gamma=(km/e\mcE)< \gamma_{\text{cr}}\equiv 1$
can become real if particle and anti-particle separate from one
another sufficiently far to acquire enough electrostatic energy. In
all cases discussed below, the electrostatic energy becoming equal to
$2m$, or $\gamma\to1$, always characterizes the onset of critical
Schwinger pair production, and a scaling behavior $\text{Im } \Gamma
\stackrel{?}{\sim} (1-\gamma^2)^\beta$ with some critical exponent
$\beta$ seems already suggestive at this stage.

An important difference to standard critical phenomena of the type
mentioned in the beginning is the occurrence of an explicit finite
mass scale: the electron mass. While universality arising near
continuous phase transitions is related to a diverging correlation
length, i.e., long-range interactions mediated by an excitation
becoming exactly massless at the critical point, the electron mass
remains as a finite scale in QED. This prevents us from
  associating the critical point with the notion of scale invariance
  and self-similarity in a straightforward way. We find that this
leads to a reduced degree of universality, implying that critical pair
production will not be characterized by a universal scaling law or
exponent, but rather by a set of scaling laws for different
large-scale properties of the spatial electric field profile. Still a
rather large degree of universality, i.e., independence of the
microscopic profile details, remains, such that electric fields fall
into universality classes of field profiles.

In the present work, we confine ourselves to simple unidirectional
electric fields that vary only in one spatial coordinate, which also
specifies the direction of the field. More precisely, we assume that
the $x$ component of the electric field can be written as
$E(x)=\mcE f'(u)$, where the potential function $f$ is antisymmetric,
monotonic and normalized such that $\max f=1$, and $u=kx$ is a
dimensionless coordinate with $1/k$ being a suitable length scale of the
spatial profile. With this restricted class of fields we avoid
pathological cases where large microscopic details could dominate the
pair production process. The latter type of fields would
  require a case by case study along the lines of fields with compact
  support included below, possibly accompanied by interference effects
  \cite{Dumlu:2011cc}. Still, the present class of fields is sufficiently
general to illustrate aspects of universality and gives access to a
variety of interesting universality classes.

We begin with the worldline representation of the effective action of
scalar QED in an external field \cite{Schmidt:1993rk} (for the
  following discussion of universality, the difference to spinor QED
  consists only in an irrelevant prefactor \cite{Dunne:2005sx})
\begin{equation}
\Gamma[A]=-\int_0^\infty \frac{ds}{s} e^{-im^2 s}\int_{x(s)=x(0)}\mathcal{D}x e^{i \int_0^s d\sigma \left( \frac{\dot{x}^2}{4} - e A\cdot \dot{x}\right)},
\label{eq:WLRep}
\end{equation}
where the path integral can be interpreted as an average over all
trajectories of electron fluctuations within the background field
$A$. Though the electron mass $m$ explicitly sets a scale, effectively
constraining the (proper-)time $s$ available for the fluctuations, the
free path integral has a Gau\ss{}ian velocity distribution such that
the ensemble contains paths of arbitrary length scale
\cite{Gies:2001zp}. This is the origin of universality for
localized fields, as the near critical regime is dominated by the
trajectories of largest relevant extent which become less and less
sensitive to the microscopic details of the background field.

In the following, we study universality in the weak-field regime,
\begin{equation}
\left(\frac{e\mcE}{m^2}\right)^2 \ll 1-\gamma^2 \ll 1.
\label{eq:scregime}
\end{equation} 
Although this prevents us from going all the way to $\gamma=1$, it is
experimentally relevant given the large value of the critical field
strength $E_{c}=\frac{m^2}{e}$. This is precisely the regime, where
the semiclassical approximation of the path integral as well as the
propertime integral in \Eqref{eq:WLRep} become exact. In this \textit{semiclassical critical} limit,
the path integral is dominated by the stationary points of the
worldline action: the worldline instantons
\cite{Affleck:1981bma,Dunne:2005sx,Dunne:2006st,Dunne:2006ur,Dietrich:2007vw,Gordon:2014aba,Strobel:2013vza}
which in general can be complex stationary paths \cite{Dumlu:2011cc}. Up to finite prefactors, the order parameter
  for pair production near semiclassical criticality scales as
  \cite{Dunne:2006st}
\begin{equation}
\text{Im }\Gamma\sim\frac{\exp\left(-\frac{\pi m^2}{eE}g(\gamma^2)\right)}{(\gamma^2g)'\sqrt{(\gamma^2g)''}},\quad (\dots)'\equiv \frac{d}{d \gamma^2} (\dots),
\label{ImGamma}
\end{equation}
where the field dependence is contained in a single function related
to the worldline instanton action
\begin{equation}
g(\gamma^2)=\frac{1}{\gamma^2}\frac{4}{\pi}\int\limits_0^{u_\gamma}\ud
u\sqrt{\gamma^2-f^2} .\label{gfunction}
\end{equation}
Here, $\pm u_\gamma$ correspond to the semiclassical turning points
defined by $f(u_\gamma)=\gamma$ (because of the anticipated
antisymmetry of $f(u)$, it suffices to consider $u>0$ here and in the
following). Heuristically, these turning points correspond to those
points, where a separated virtual pair has acquired sufficient
electrostatic energy to become real.

\Eqref{ImGamma} has the standard semiclassical form of an exponential
tunneling amplitude arising from the action along a classical path,
and a prefactor from the fluctuations about the classical path. The
order parameter $\text{Im }\Gamma$ vanishes if the prefactor vanishes,
i.e., $g', g''$ diverge, or if the exponent $\sim g$ diverges.

Universality becomes already apparent from the dependence of $g$ on
the potential function $f$ in \Eqref{gfunction}: \textit{any field
  strength profile that leads to the same divergence structure of $g$
  or its derivatives for $\gamma\to1$ belongs to the same universality
  class.} In the limit $\gamma\to1$, the turning
point $u_\gamma$ approaches the point $u_0$ where the electric field
vanishes and $f$ attains its maximum
\begin{equation}
f(u_\gamma)=\gamma\to1=\max f=:f(u_0) . \label{eq:maxcond} 
\end{equation}
Substituting $\sin\theta=f(u)/\gamma$, we find
\begin{equation}\label{eq:gprime}
(\gamma^2 g(\gamma^2))'=\frac{2}{\pi}\int\limits_0^{\pi/2}\ud
\theta\frac{1}{f'} \;, 
\end{equation}
demonstrating that the divergence of $g'$ comes from the region
close to zero field strength and the maximum of the potential,
$f'\to0$. Actually, while $g''$ always diverges, $g'$ can be finite
for certain compact fields (see below). Further, for fields vanishing
asymptotically, $u_0\to\infty$, also $g$ can diverge; otherwise, e.g.,
for fields with compact support, $g$ is finite for regular fields and
the tunneling amplitude (i.e. the exponent in \Eqref{ImGamma}) cannot
contribute to criticality.

\begin{figure}[t]
\includegraphics[width=0.45\textwidth]{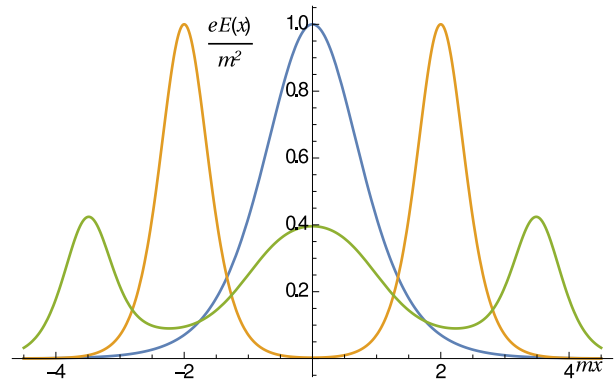}
\caption{Various examples for critical field profiles with
  exponent $\beta=\frac{5}{4}$. The onset of criticality is determined
  by the asymptotic behavior (exponential in these cases). The
  critical scaling law \Eqref{eq:scaling1} is independent of the local
  details of the field profiles.}
\label{fig}
\end{figure}

The resulting scaling laws can analytically be extracted by expanding
$f$ near the leading-order divergence of $1/f'$. Let us consider
several paradigmatic examples, starting with localized fields that
decay asymptotically with a power, $E(x)\sim \mcE \frac{c}{(kx)^p}$ as $x\to\infty$, with some
constant $c$. Then, $f\approx 1-\frac{c}{p-1}\frac{1}{u^{p-1}}$ and
$u_0\to \infty$. Depending on the power $p$, three different cases
occur: (I) for $p>3$, the function $g$ stays finite and the scaling law
arises purely from the fluctuation prefactor, yielding a standard
power-law
\begin{equation}
\text{Im }\Gamma\sim(1-\gamma^2)^\critexp, \quad \critexp=\frac{5p+1}{4(p-1)}.
\label{eq:scaling1}
\end{equation}
We emphasize that all field profiles with the same power-law decay
exhibit the same universal critical scaling independently of the
details of the profile at finite $x$ (at least within the class of fields
specified above). Equation \eqref{eq:scaling1} also includes
exponentially decaying fields: in the limit $p\to\infty$, we discover
a unique exponent $\critexp=\frac{5}{4}$.  This agrees, for instance,
with the exact result \cite{Nikishov:1970br} for the $\text{sech}^2
kx$ profile which in the regime \eqref{eq:scregime} reads
\begin{equation}
\text{Im }\Gamma=\frac{L^2T m^3}{2(2\pi)^3 }
\left(\frac{e\mcE}{m^2}\right)^{3/2}(1-\gamma^2)^{5/4}e^{-\frac{2\pi
    m^2}{e\mcE}}.
\end{equation}
We emphasize that the exponent $\critexp=\frac{5}{4}$ also holds for other
exponentially localized fields, different examples are shown in Fig.~\ref{fig}. 

(II) The powerlaw decay $p=3$ is special, since the function $g$
itself diverges logarithmically which -- upon insertion into
\Eqref{ImGamma} -- produces a field-dependent power in addition to
$\critexp=2$,
\begin{equation}
\text{Im }\Gamma\sim(1-\gamma^2)^{2\left(1+\sqrt{c}\frac{m^2}{e\mcE}\right)} \;.\label{Lorentz}
\end{equation}
Since $\frac{e\mcE}{m^2}\ll 1$, cf. \Eqref{eq:scregime}, this
field-dependent part dominates the exponent, indicating the approach to exponential
scaling.

(III) The latter becomes manifest for a field decaying with $1<p<3$,
since the instanton action $\sim g$ diverges as a power near criticality,
resulting in the scaling law
\begin{equation}
\text{Im }\Gamma\sim (1-\gamma^2)^\beta \exp\left(-\frac{\pi m^2}{e\mcE} \frac{C}{(1-\gamma^2)^\lambda} \right),
\label{eq:essscal}
\end{equation}
where the \textit{essential} exponent $\lambda$ and the constant $C$,
\be
\lambda=\frac{3-p}{2(p-1)}, \quad C=\frac{2}{\pi c}\bigg(\frac{2c}{p-1}\bigg)^\frac{p}{p-1}B\Big(\frac{3}{2},\frac{3-p}{2(p-1)}\Big),
\ee
are both universal. (For $p<5/3$, the exponent in \Eqref{eq:essscal}
can acquire universal subleading singularities
e.g. $1/(1-\gamma^2)^{\lambda-1}$ or $\ln(1-\gamma^2)$.)  In critical
phenomena, a scaling of this type is known as essential scaling or BKT
(or Miransky) scaling \cite{Berezinsky:1970fr}. It is known to occur
in a wide range of systems, in particular those exhibiting a
transition from a conformal to a non-conformal phase
\cite{Miransky:1996pd}. While our scaling law includes the BKT-scaling
law with exponent $\lambda=\frac{1}{2}$ for an electric field decaying
with power $p=2$, any essential exponent $\lambda>0$ can be realized
for appropriate decay powers $p$. Equation \eqref{eq:essscal} also has
a universal powerlaw prefactor which is reminiscent to the many-flavor
phase transition in gauge theories \cite{Braun:2009ns}. We also
observe that $\lambda$ diverges for $p\to1$ where the electrostatic
energy receives dominant contributions from long-range
fluctuations. Essential scaling of critical pair production hence is
obviously related to a dominance of electron-positron fluctuations at
the largest length scales.
 
Let us now turn to electric fields of compact support in $x$
direction. Within the class of fields considered here, this implies
that the potential function $f(u)$ attains its maximum at a finite
value $u_0$. Correspondingly, $E(x)=0$ for $|x|>u_0/k$. The worldline
action \eqref{gfunction} cannot become singular in this case, so the
scaling law is solely determined by the fluctuation prefactor and thus
by the way in which $f'(u)$ approaches zero for $u\to u_0$,
cf. \eqref{eq:gprime}. Let us assume that the electric field drops to
zero as $E(x)\sim (u_0/k - x)^n$. For $n>1$, the order parameter
satisfies power-law scaling \eqref{eq:scaling1} with exponent
\begin{equation}\label{scaling-n}
\critexp=\frac{5n-1}{4(n+1)}.
\end{equation}
It is interesting to see that the universal exponent for exponential
decay $\critexp=\frac{5}{4}$ is rediscovered in the limit
$n\to\infty$. Note also that \eqref{scaling-n} can be obtained by
replacing $p\to-n$ in \eqref{eq:scaling1}.

Another special case is $n=1$, where we find a logarithmic divergence
in the prefactor, $g'\sim -\ln (1-\gamma^2)$, such that the scaling
law receives log corrections
\begin{equation}
\text{Im }\Gamma\sim\frac{(1-\gamma^2)^{1/2}}{-\ln(1-\gamma^2)}. \label{eq:scaling3}
\end{equation}
Again, this has an analog in statistical physics, as log-corrections
are known to arise in cases where marginal operators contribute to
criticality \cite{Wegner:1972}, such as in the $2d$ 4-state
  Potts model \cite{Cardy:1980}. For $n<1$ only $g''$ diverges and we
  again find power-law scaling \eqref{eq:scaling1} with
\begin{equation}
\critexp=\frac{3n+1}{4(n+1)} \;.
\end{equation}
In the limit $n\to0$, the electric field becomes step-like with exponent $\critexp=\frac{1}{4}$.

The diversity of universality classes given above can be covered in a
unified description using an implicit definition of the electric field
in terms of differential equations. One such possible definition is
\begin{equation}
\label{implicit-def}
f'=(1-f^2)^b ,
\end{equation}
with the implicit solution $u=f\,{_2}F_1(1/2,b,3/2,f^2)$ in terms of a
hypergeometric function. The case, $b>1$ covers all examples of
asymptotically vanishing fields given above with power $p=b/(b-1)$,
and $b<1$ corresponds to fields with compact support with
$n=b/(1-b)$. The cases $b=1/2, 1$ and $3/2$ correspond to to the
fields $E\sim\cos kx$, $\text{sech}^2kx$ and $(1+kx^2)^{-3/2}$,
respectively, studied explicitly in \cite{Dunne:2006st}. We have
$g={_2}F_1(1/2,b,2,\gamma^2)$, and thus
$g''\sim(1-\gamma^2)^{-(b+1/2)}$, cf. \cite{dlmf}, which agrees with
all the different scalings above.

This unified analysis also verifies the general trend of the relation
between field profile and scaling: field profiles which are more
spread out exhibit a steeper scaling. So, the order parameter
$\text{Im }\Gamma$ vanishes faster for an asymptotically decaying
field than for a compact field. This is in agreement with the
Euclidean worldline picture, since the contributions from large-scale
fluctuations (large propertimes) are suppressed by the electron mass
scale.

The similarity of critical Schwinger pair production to critical
phenomena discovered in this work appears to call for a
renormalization group description. It is conceivable that the critical
point corresponds to a fixed point of a suitable coarse-graining
procedure involving the worldlines, the background field or both. Such
a description would be rewarding as it could give access to potential
further aspects of criticality such as (hyper-)scaling relations.

Our results can straightforwardly be generalized to field
  profiles with only asymptotic symmetry as well as to an arbitrary
number of translation invariant transversal directions ($y,z,t$ in the
present work). As the latter only influences the propertime integrand, $1/s
\to 1/s^{\frac{d-2}{2}}$, the order parameter receives additional
scaling factors according to $\text{Im
}\Gamma_d\sim(\gamma^2g)'^{\frac{4-d}{2}}\text{Im }\Gamma_4$ with
corresponding consequences for the scaling exponents.

From the underlying picture in terms of virtual fluctuations needing
to acquire sufficient energy to become real, we expect
that our results analogously persist also for static fields that
are localized in more than one space dimension, even though the
analysis can become rather involved for non-unidirectional fields
\cite{Dunne:2006ur}. For time-dependencies slower than the Compton
scale, the existence of a critical point $\gamma_{\text{cr}}\simeq 1$
has been observed in \cite{Hebenstreit:2011wk,Schneider:2014mla}. We
expect though a qualitative change for fields varying rapidly in
time. Indeed, there are no critical points for fields depending on
lightfront time $t+x$ \cite{Tomaras:2001vs,Ilderton:2014mla}. As
soon as the fields vary in time, pair production does not have to rely
only on (instantaneous) electrostatic energy, but can also be
supported by finite (multi-)photon energies of the varying
field. Hence, we expect the critical point in the spatial adiabaticity
parameter to shift to values larger than $\gamma_{\text{cr}}=1$ for
increasing time variations. This claim is supported by a recent paper
\cite{Ilderton:2015qda} which shows that for electric fields depending
on a coordinate $q$ that interpolates continuously between $x$, $t+x$
and $t$, the critical point increases from $\gamma_\text{cr}=1$ at
$q=x$ to infinity at $q=t+x$ (i.e. effectively vanishes);  for
timelike $q$ there is no critical point.

For rapidly varying fields, pair production via multi-photon effects
may remove any singularity associated with criticality. The would-be
critical point due to the spatial profile may then still be visible as
a cross-over in the production rate.

We emphasize that our results for universality hold in the regime
defined by \Eqref{eq:scregime}. We expect analogous features also in
the deeply critical regime where $ 1-\gamma^2 \ll
\left(\frac{e\mcE}{m^2}\right)^2 \ll 1$, even though the values of the
exponents might differ. For instance, the exact solution for the
$\mathrm{sech}^2 kx$ case scales differently in this regime, $\text{Im
}\Gamma \sim(1-\gamma^2)^3$ \cite{Nikishov:1970br}. Determining the
degree of universality in this regime remains an interesting
problem. As the electron mass scale is less dominant,
  universality could even be substantially enhanced.

Radiative corrections will also take a subleading quantitative
influence on our results. E.g., the two-loop correction includes
mass-shift effects \cite{Ritus:1975cf,Dunne:2002qg}; a resummation 
could also account for a production into a positronium bound state. As
these effects modify the invariant mass of the final state, they may
primarily lead to a deviation of the critical point
$\gamma_{\text{cr}}\simeq 1$ but could preserve the universal critical
exponents.

In summary, we have discovered an analogy between Schwinger pair
production and continuous phase transitions. This analogy is
quantitatively manifest in universal scaling laws for the onset of
pair production in spatially inhomogeneous electric backgrounds. The
scaling laws show a high degree of universality as the corresponding
critical exponents only depend on the large-scale properties of the
background (for monotonic potentials) but become insensitive to the
microscopic details. Hence, localized electric backgrounds fall into
universality classes each giving rise to a characteristic scaling
law. As a particularly fascinating aspect, we discovered universality
classes covering essentially all types of scaling laws familiar from
continuous phase transitions.

\acknowledgments

We thank Gerald Dunne, Anton Ilderton, Felix Karbstein, and Ralf
Sch\"utzhold for interesting discussions. G.T. thanks TPI, FSU Jena, and
HI Jena for hospitality during a research visit. We acknowledge
support by the DFG under grants No. GRK1523/2, and SFB-TR18 (H.G.),
and the Swedish Research Council, contract 2011-4221 (P.I.:
A.~Ilderton), (G.T).

\end{document}